\documentclass[epj,referee]{svjour}
\usepackage[dvips]{graphicx}
\usepackage{amsmath}
\def\EQ{\begin{equation}}
\def\EN{\end{equation}}
\def\EQA{\begin{eqnarray}}
\def\ENA{\end{eqnarray}}

\def\xx{{\bf x}}

\begin{document}

\title{The turbulent dynamo as an instability in a noisy medium}
\author{N. Leprovost and B. Dubrulle}

\institute{ Groupe Instabilit\'e et Turbulence,
\\CEA/DSM/DRECAM/SPEC and CNRS. URA 2464 F-91191 Gif sur Yvette Cedex, France
\\ \email{nicolas.leprovost@cea.fr}}

\date{Received: date / Revised version: date}

\abstract{We study an example of instability in presence of a multiplicative noise, namely the spontaneous generation of a magnetic field in a turbulent medium. This so-called turbulent dynamo problem remains challenging, experimentally and theoretically. In this field, the prevailing theory is the Mean-Field Dynamo \cite{Krause80} where the dynamo effect is monitored by the mean magnetic field (other possible choices would be the energy, flux,...). In recent years, it has been shown on stochastic oscillators that this type of approach could be misleading. In this paper, we develop a stochastic description of the turbulent dynamo effect which permits us to define unambiguously a threshold for the dynamo effect, namely by globally analyzing the probability density function of the magnetic field instead of a given moment. }


\maketitle

\vspace{0.1cm}

Classical stability analysis are usually performed in systems were the control parameter is a non-fluctuating quantity, e.g. for laminar flows in hydrodynamics. When the instability occurs in a random system (e.g. a turbulent medium), resulting fluctuation of the control parameter, or multiplicative noise, may generate several surprising effects that have been studied in a variety of systems. The possibility of stabilization by noise has first been evidenced on a Duffing oscillator \cite{Bourret73} , where the solution $x(t) = 0$ is stable for values of the control parameter above the deterministic threshold. This stabilization is generic for weak intensities of the noise. For stronger intensity, however, noise induced transition may also arise in this system \cite{Mallick03}. In the case of a parametric instability, it has been showed experimentally \cite{Berthet03} that the instability is sensitive to the bifurcation nature. In the supercritical case, oscillatory bursts (corresponding to a signal with a vanishing mean but a most probable value equal to zero) appear first when the control parameter is increased and are then replaced by a state where the most probable value is no more equal to zero. On the contrary, in the subcritical case, there is coexistence between these two states. This illustrates a central difficulty of instability in presence of multiplicative noise associated with an ambiguity regarding the threshold value, which depends on the definition of the order parameter \cite{Strato67}.

The observations and techniques developed in these simple systems may be used to shed new light on some recent issues associated with the dynamo effect, the process of magnetic field generation through the movement of an electrically conducting medium. In this case, the instability results from a competition between amplification of a seed magnetic field via stretching and folding, and magnetic field damping through diffusion. In a laminar fluid, it is controlled by a dimensionless number, the magnetic Reynolds number ($Rm$), which must exceed some critical value $Rm_c$ for the instability to operate. In a turbulent medium, velocity fluctuations induce fluctuation of the control parameter, making the turbulent dynamo problem similar to an instability in the presence of multiplicative noise. In that respect, recent numerical findings such as observed in \cite{Sweet01} may find a natural explanation. In their work, the authors observed short intermittent bursts of magnetic activity separated by relatively long periods, increasing towards the bifurcation threshold. This feature could be explained in terms of a supercritical instability in presence of multiplicative noise since in this case, the bifurcated state is generally composed of oscillatory bursts. More generally, the multiplicative noise paradigm could turn useful to interprete the outcome of recent experiments involving liquid metals. Among the various operating experiments, a clear distinction appears between set up with constrained or unconstrained geometry. In the former case \cite{Gailitis00}, the fluctuation level is very weak. The velocity field is then very close to its laminar (mean) value. In these experiments, dynamos have been observed, at critical magnetic Reynolds number comparable to the theoretical value. In contrast, unconstrained experiments \cite{Peffley00} are characterized by a large fluctuation level (as high as 50 per cent). A surrogate laminar $Rm_c$ can then be computed, using the mean velocity field as an input \cite{Marie03} but it is not clear whether it will correspond to the actual dynamo threshold, owing to the influence of turbulent fluctuations.

In this paper, we investigate this issue by techniques developed to study the Duffing oscillator and using a stochastic description of small-scale turbulent motions. This subject has been pioneered by Kazantsev \cite{Kazantsev68}, Parker \cite{Parker70} and Kraichnan \cite{Kraichnan67}, and further developed by the Russian school \cite{Zeldovich84}. It has recently been the subject of a renewed interest, in the framework of anomalous scaling and intermittency \cite{FGV01}, or computation of turbulent transport coefficients and probability density functions (PDF) \cite{Boldyrev01}.

The dynamic of the magnetic field ${\bf B}$ in an infinite conducting medium of diffusivity $\eta$ and velocity ${\bf V}$, is governed by the induction equation:
\EQ
\partial_t B_i = - V_k \partial_k B_i + B_k \partial_k V_i + \eta \partial_k \partial_k B_i \; ,
\label{induction}
\EN
with control parameter built using typical velocity and scale as $Rm=LV/\eta$.
We decompose the velocity field into a mean part $\bar{V}_i$ and a fluctuating part $v_i$. In most laboratory experiments, the mean part is provided by the forcing. As such, it is generally composed of large scales, while the fluctuating part collects all short time scale, small-scale movements. In this regard, it is natural to consider the fluctuating part of the velocity as a noise, to be prescribed or computed in a physically plausible manner. The simplest, most widely used shape is the Gaussian, delta-correlated fluctuations, the so-called \lq\lq Kraichnan's ensemble\rq\rq:
\EQ
\label{Kraichnan}
\langle v_i(\xx,t) v_j(\xx',t') \rangle = 2 G_{ij}(\xx,\xx') \delta(t-t')
\EN
Equation (\ref{induction}) then takes the shape of a stochastic partial differential equation for $B$. In that respect, we note that the induction equation is {\sl linear} and does not include explicit back reaction term allowing saturation of any potential growth in the dynamo regime. This back reaction is provided through the velocity which is subject to the Lorentz-Force, a quadratic form of $B$. It is usually ignored in the so-called kinematic regime. However, for reasons which will become clearer later, we prefer to work with a modified induction equation, so as to model this non-linear back reaction.\

Indeed, the study of the induction equation alone could be misleading when looking at the threshold of the dynamo effect (it leads to a threshold value dependent on the considered moment). This is symptomatic of the fact that the dynamo problem is {\em nonlinear} through the Navier-Stokes equation. A practical way to include the effect of the Lorentz force at the onset of the nonlinear regime is to add a saturating term in the induction equation. Symmetry considerations then favor a term like $- c B^2 B_i$. In some sense, this modification is akin to an amplitude equation, and the cubic shape for the non-linear term could be viewed as the only one allowed by the symmetries. Such a procedure has been validated by \cite{Petrelis02} in the case of the saturation of a Ponomarenko dynamo. Such a cubic form has also been evidenced by Boldyrev \cite{Boldyrev01} by assuming the equality of viscous and dynamical stresses in the Navier-Stokes equation at the onset of backreaction. In the sequel, we show that the precise form of the nonlinear term does not affect the threshold value, which only depends on the behaviour for $\vert B\vert\to 0$. This is similar to the case of the Duffing oscillator where the threshold is given by the Lyapunov exponent of the {\em linearized} oscillator \cite{Mallick03}.

A further difficulty is associated with the presence of the diffusive terms. Their physical influence is to cut-out magnetic field Fourier components with wave numbers initially oriented in the contracting directions (stretching directions for wave numbers). Through the divergence free condition, this favors magnetic field components pointing towards contracting directions, thus counteracting the initial effect of magnetic growth along stretching direction. The balance between these two effects yields the dynamo threshold (\cite{Chertkov99}). The direct consideration of diffusive terms in the stochastic formulation requires functional derivative and integration, hindering simple analytical description. We propose to model them partially through an additional "molecular" homogeneous noise $\xi_i(t)$, superposed to and decorrelated from the velocity fluctuation \cite{Pavel02}, with correlation function $<\xi_i(t)\xi_j(t')>=2\eta\delta_{ij}\delta(t-t')$. In the sequel, we show that this choice provides some sort of saturation for the moments of various order, similar to the role of a viscosity. Damping of magnetic field fluctuations, however, is not properly taken account by this model. \

Using standard techniques \cite{ZinnJustin,Boldyrev01}, one can then derive the evolution equation for $P({\bf B},\xx,t)$, the probability of having the field ${\bf B}$ at point $\xx$ and time $t$ (we assume an homogeneous turbulence for simplicity):
\EQA
\label{FP}
\partial_t P &=& - \bar{V}_k \partial_k P - (\partial_k \bar{V}_i) \partial_{B_i} [B_k P] + \partial_k [\beta_{kl} \partial_l P] \\ \nonumber
&+& c \partial_{B_i}[B^2 B_i P] + 2 \partial_{B_i} [B_k \alpha_{lik} \partial_l P] \\ \nonumber
&+& \mu_{ijkl} \partial_{B_i} [B_j \partial_{B_k} (B_l P)] \; ,
\ENA
with the following turbulent tensors:
\EQA
\label{tenseurs}
\beta_{kl} &=& \langle v_k v_l \rangle+\eta\delta_{kl} , \quad \alpha_{ijk} = \langle v_i \partial_k v_j \rangle \\ \nonumber
&& \text{and} \quad \mu_{ijkl} = \langle \partial_j v_i \partial_l v_k \rangle \; .
\ENA

The physical meaning of these tensors can be found by analogy with the \lq\lq Mean-Field Dynamo theory\rq\rq \cite{Krause80,Moffatt78}. Indeed, consider the equation for the evolution of the mean field, obtained by multiplication of equation (\ref{FP}) by $B^i$ and integration:
\EQA
\label{MFE}
\nonumber
\partial_t \langle B_i \rangle &=& - \bar{V}_k \partial_k \langle B^i \rangle + (\partial_k \bar{V}_i) \langle B_k \rangle - 2 \alpha_{kil} \partial_k \langle B_l \rangle \\
&+& \beta_{kl} \partial_k \partial_l \langle B_i \rangle - c \langle B^2 B_i \rangle.
\ENA
This equation resembles the classical Mean Field Equation of dynamo theory, with generalized (anisotropic) \lq\lq $\alpha$\rq\rq and \lq\lq$\beta$\rq\rq. The first effect leads to a large scale instability for the mean-field, while the second one is akin to a turbulent diffusivity. A few remarks are in order at this point: i) our mean field equation has been derived without assumption of scale separation. ii) The tensor $\mu$ does not appear at this level. In the sequel, it will be shown to govern the stochastic dynamo transition.\

For this, we need to identify the threshold as a function of the noise properties. Here, we follow an idea by Mallick and Marcq \cite{Mallick03}, and focus on the properties of the stationary PDF of the system. Indeed, below the transition, the only stable state is $B=0$ and the PDF should be a Dirac delta function. Above the transition, other equilibrium states are possible, with non zero magnetic field. However, in the general case, it is not possible to find analytical solution for the equation (\ref{FP}). We thus resort to the following mean field argument. Changing variable $B_i=B e_i$ where $e$ is a unit vector (and can be characterized by $d-1$ angular variables), we can get an equation for $P(B,e_i, x) = J P(B_i)$ where $J=B^{d-1}$ is the Jacobian of the transformation. We now assume that there is an uncoupling for $P$ as $P(B,e_i)=P(B)G(e_i,x)$, and perform an average over the angular variables, to find a closed equation for $P(B)$. In some sense, this can be regarded as a kind of angle/action variable separation, with average over the fast variables. The final equation for $P(B)$ becomes:
\EQ
\label{equanorm}
\frac{\partial P}{\partial t} = a \frac{\partial}{\partial B}[B \frac{\partial}{\partial B} (B P)] - b \frac{\partial}{\partial B} (B P) + c \frac{\partial}{\partial B}(B^3 P)
\EN
where the coefficients are given by averages over the position and the angular variables $\langle \bullet \rangle_{\phi} = \int \bullet \; G({\bf e},\xx) \; d\xx d{\bf e}$:
\EQA
a &=& \langle \mu_{ijkl} e_i e_j e_k e_l \rangle_{\phi} \\ \nonumber
b &=& \langle \partial_k \bar{V}_i e_i e_k \rangle_{\phi} + \langle \mu_{ijkl} (\Delta_{ik} e_j e_l + \Delta_{kj} e_i e_l) \rangle_{\phi}
\ENA
where we used $\Delta_{ij}=\partial_{n_i}(n_j)= \delta_{ij} - e_i e_j $ an \lq\lq angular Dirac tensor\rq\rq. One can notice that these coefficients only explicitly involve the tensor $\mu$. Nevertheless, one must keep in mind that the tensor $\alpha$ and $\beta$ enter these expressions by mean of the angular distribution $G(e_i,x)$, whose expression involves this two tensors in the general case.

An obvious stationary solution of (\ref{equanorm}) is a Dirac function, representing a solution with vanishing magnetic field. Another stationary solution can be found by setting $\partial_t P = 0$ in (\ref{equanorm}), with solution:
\EQ
\label{solution}
P(B) = \frac{1}{Z} B^{b/a-1} \exp\Bigl[-\frac{c}{2 a} B^2\Bigr]
\EN
where $Z$ is a normalization constant. This solution can represent a meaningful probability density function only if it can be normalized. This remark provides us with a bifurcation threshold: there is dynamo whenever (\ref{solution}) is integrable, i.e., when solution other than vanishing magnetic field are possible.

\begin{figure}[h]
\begin{center}
\includegraphics[scale=0.48,clip]{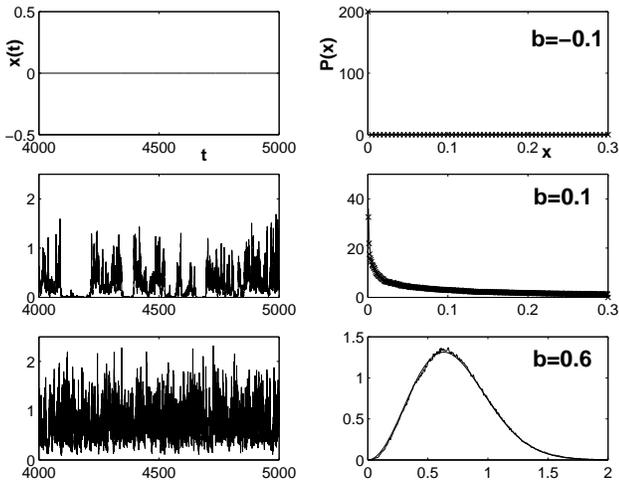}
\caption{\label{Bifurcation} Result of the surrogate 1D model: $\partial_t x = (b+\xi(t)) x - \gamma x^3$ with $\langle \xi(t)\xi(t') \rangle = 2a\delta(t-t')$. On the left side we show time series for $a = 0.2$, $\gamma=1$ and 3 different values of the parameter $b$. On the right side, the corresponding PDF and the theoretical curve corresponding to equation (\ref{solution}).}
\end{center}
\end{figure}

Condition of integrability at infinity of (\ref{solution}) requires $a$ be positive. This illustrates the importance of the non-linear term which is essential to ensure vanishing of the probability density at infinity. Condition of integrability near zero requires $b/a$ be positive. This leads us to identify a necessary and sufficient condition for existence of a stationary dynamo as
\EQ
\label{dynamo}
a > 0 \quad \mathrm{and} \quad \frac{b}{a} > 0 \qquad{DYNAMO}
\EN
In some sense, this bifurcation (I) is obtained using the mean field as control parameter. Another bifurcation threshold can be defined using the most probable field as control parameter.
Indeed, an elementary calculation shows that the condition for a maximum in the PDF is $b > a$. Therefore, the bifurcation threshold (II) with the most probable field as control parameter is defined by $b=a$. This difference may have some relevance when analyzing real data from experiment. To illustrate this, we performed simulations of the 1-D version of our non-linear stochastic system. It may be checked that the stationary PDF in this case is exactly given by eq. (\ref{solution}). Therefore, the time series and associated PDF are good illustration of the output of our 3D model, and the meaning of the two bifurcations discussed above. The time series and PDF for three different values of the control parameter are shown on figure \ref{Bifurcation}.

The simulations show that the bifurcation (I) leads to an intermittent behavior for the magnetic field reminiscent of the characteristic behavior of instability in presence of multiplicative noise: typically, the magnetic energy exhibits bursts separated by long quiescent (zero magnetic energy) period. Equation (\ref{solution}) cannot rigorously capture these intermittent states. However, two facts are very suggestive of such a type of bifurcation in our solution: (a) the distribution looks like a pure fluctuation distribution, with ill-defined mean value (b) the scaling for $\| B \| \ll 1$ (magnetic energy) is the same as that of \cite{Sweet01}: $P(\| B \|) = \| B \|^\gamma$ with $\gamma=b/D-1$, where $D$ is a \lq\lq diffusion coefficient\rq\rq for the finite-time Lyapunov exponent. On the contrary, the bifurcation (II) is quite different in nature because of a well defined mean value for the magnetic field and fluctuations around this mean.\
Note also that if we consider the dynamo instability in absence of noise, $a=0$, the two bifurcation threshold collapse. As stated by \cite{Sweet01}, the bifurcation corresponding to $b > 0$ may be difficult to observe in real experiments because, under the threshold, the presence of the Earth external magnetic field always gives rise to magnetic fluctuations qualitatively similar to that above the threshold (magnetic bursts separated by quiescent period). This effect can be taken into account by adding an additive noise to equation (\ref{induction}).

In the theory of dynamical systems stability, the instability criterion is usually associated with the existence of a positive Lyapunov exponent for the growth of the system energy: $\lim \ln B^2 / 2t = \lim \ln B /t$. It is possible to find this exponent, by multiplying equation (\ref{equanorm}) by $\ln B$ and integrating with respect to $B$. This yields $\partial_t \langle \ln B \rangle = b \,$, meaning that the Lyapunov exponent in our system is equal to $b$. The two instability criteria (existence of a normalizable solution or a positive Lyapunov exponent) are therefore identical provided $a > 0$, a necessary condition for integrability of the PDF at infinity.

It is now interesting to discuss qualitatively the meaning of our main result (\ref{dynamo}). It is possible to show that for isotropic or axisymmetric velocity fluctuations, the coefficient $a$ is positive. So we suspect that the main condition for existence of a dynamo is positivity of $b$. Therefore, in the limit of zero noise, the term proportional to $\mu$ is negligible and the dynamo threshold is only determined by the condition $\langle \partial_k \bar{V}_i e_i e_k \rangle_{\phi} > 0$. Since the magnetic field will mainly grow in the direction given by the largest eigenvalue of $S_{ij}=\partial_j \bar V_i$ and that the molecular diffusivity will tend to orientate the magnetic field along contracting direction, the dynamo threshold will be the same as in the deterministic case. Consider now a situation where you increase the noise level. Two different influences on the sign of $b$ then result: one through the factor proportional by $\mu$. According to the sign of this factor, it can therefore favor or hinder the dynamo. Another less obvious influence is through vector orientation. Indeed, noise changes the distribution of magnetic field orientation. For example, if noise induces a flat distribution for $e_i$, then $\langle \partial_k \bar{V}_i e_i e_k \rangle_{\phi} = S_{ii} =0$. Even more dramatic results can be obtained if the noise tends to align the vector along a direction of negative eigenvalue for $S_{ij}$, since in that case the factor $\langle \partial_k \bar{V}_i e_i e_k \rangle_{\phi}$ becomes negative, decreasing the dynamo threshold. From these two examples, one sees that in the presence of noise, the dynamo threshold can be completely disconnected from the dynamo threshold in the deterministic system ! Some caution is therefore in order when designing an experiment based on mean field measurements.\

Our approach gives a quantitative criterion on the dynamo threshold (namely $b > 0$). However, its practical implementation requires the measure on the angular and position variables $G({\bf x}, {\bf n}, t)$ and the average of the tensor $\mu$ with this measure. One can obtain the measure by integrating equation (\ref{FP}) with respect to $B$. Unfortunately the equation for $G$ can not be solved in the general case and particular types of turbulence statistics have to be considered (isotropic, axisymmetric, etc...). Work is under progress to determine the angular measure in these simple cases. It is however interesting to note that this measure explicitly involves the tensors $\alpha$ and $\beta$ defined in (\ref{tenseurs}) and appearing in the Mean Field Equations (\ref{MFE}). In that sense, the dynamo threshold depends on these tensors, albeit in a less explicit way than in the Mean Field Equations (MFE). It would therefore be interesting to confront threshold derived from (MFE), which are $\mu$ independent, and from our theory, to see what kind of error in the threshold determination one can expect by using MFE instead of the true, non-perturbative theory.

{\bf Acknowledgments}\
We thank the Programme national de Plan\'etologie, and the GDR Turbulence and GDR Dynamo for moral and financial support and F. Daviaud, K. Mallick and P. Marcq for discussion and comments. We also thank Y. Pomeau for stimulating the study of the intermittent dynamo and F. P\'etr\'elis for pointing out the link with the \lq\lq on-off\rq\rq intermittency.


\begin{thebibliography}{}
\bibitem{Krause80} F. Krause and K.-H. R{\"a}dler, {\it Mean field MHD and dynamo theory} (Pergamon press, 1980).
\bibitem{Bourret73} R. Bourret, U. Frisch and A. Pouquet, Physica {\bf 65}, 303 (1973); R. Graham and A. Schenzle, Phys. Rev. A {\bf 26}, 1676 (1982); M. L\"ucke and F. Schank, Phys. Rev. Lett. {\bf 54}, 1465 (1985).
\bibitem{Mallick03} K. Mallick and P. Marcq, Eur. Phys. J. B {\bf 36}, 119 (2003); {\bf 38}, 99 (2004).
\bibitem{Berthet03} S. Residori, R. Berthet, B. Roman and S. Fauve, Phys. Rev. Lett. {\bf 88}, 024502 (2002); R. Berthet {\it et al.}, Physica D {\bf 174}, 84 (2003).
\bibitem{Strato67} R. L. Stratonovich, {\it Topics in the Theory of Random Noise} (Gordon and Breach, 1967).
\bibitem{Sweet01} D. Sweet {\it et al}, Phys. Rev. E {\bf 63}, 066211 (2001).
\bibitem{Gailitis00} A. Gailitis {\it et al.}, Phys. Rev. Lett. {\bf 84}, 4365 (2000); R. Stieglitz and U. M{\"u}ller, Phys. Fluids {\bf 13}, 561 (2001).
\bibitem{Peffley00} N. L. Peffley, A. B. Cawthorne and D. P. Lathrop, Phys. Rev. E {\bf 61}, 5287 (2000); M. Bourgoin {\it et al}, Phys. Fluids {\bf 14}, 3046 (2002).
\bibitem{Marie03} L. Mari\'e {\it et al.}, Eur. Phys. J. B {\bf 33}, 469 (2003); F. Ravelet, A. Chiffaudel, F. Daviaud and J. L\'eorat, submitted to Phys. Fluids.
\bibitem{Kazantsev68} A. P. Kazantsev, Sov. Phys. JETP {\bf 26}, 1031 (1968).
\bibitem{Parker70} E. N. Parker, Astrophys. J. {\bf 162}, 665 (1970).
\bibitem{Kraichnan67} R. H. Kraichnan and S. Nagarajan, Phys. Fluids {\bf 10}, 859 (1967).
\bibitem{Zeldovich84} Ya. B. Zel'dovich, A. A. Ruzmaikin, S. A. Molchanov and D. D. Sokolov, J. Fluid Mech. {\bf 144}, 1 (1984).
\bibitem{FGV01} G. Falkovich, K. Gawedzki and M. Vergassola, Rev. Mod. Phys. {\bf 73}, 913 (2001).
\bibitem{Boldyrev01} S. Boldyrev, Astrophys. J. {\bf 562}, 1081 (2001).
\bibitem{Petrelis02} F. P\'etr\'elis, PhD Thesis {\bf Paris VI}, (2002); S. Fauve and F. Petrelis, {\it The dynamo effect}, in Peyresq Lectures on nonlinear phenomena (World scientific, 2003).
\bibitem{Pavel02} S. Pavel, S. Berloff and J. C. McWilliams, J. Phys. Oceanogr. {\bf 32}, 797 (2002).
\bibitem{Chertkov99} M. Chertkov, G. Falkovich, I. Kolokolov and M. Vergassola, Phys. Rev. Lett. {\bf 83}, 4065 (1999).
\bibitem{ZinnJustin} J. Zinn-Justin, {\it Quantum Field Theory and Critical Phenomena} (Oxford, 2002).
\bibitem{Moffatt78} H.K. Moffatt, {\it Magnetic field generation in electrically conducting fluids} (Cambridge University Press, 1978).
\end{thebibliography}
\end{document}